\documentclass{article}
\usepackage{emulateapj}
\usepackage{times}
\usepackage{epsfig}

\begin{document}

\submitted{}
\title{Astrometric Microlensing of Stars}  

\author{Martin Dominik\altaffilmark{1,2} and Kailash C. Sahu\altaffilmark{1}}
\affil{{\em e-mail}: dominik@astro.rug.nl, ksahu@stsci.edu}
\altaffiltext{1}{Space Telescope Science Institute, 3700 San Martin Drive, Baltimore, MD 21218, USA}
\altaffiltext{2}{Kapteyn Astronomical Institute, Postbus 800, 9700 AV Groningen,
The Netherlands}

\begin{abstract}
Due to dramatic improvements in the precision of astrometric measurements,
the observation of light centroid shifts in observed stars due to
intervening massive compact objects (`astrometric microlensing') will become
possible in the near future. Upcoming space missions, such as SIM and GAIA,
will provide measurements with an accuracy of 4--60 $\mu\mbox{as}$ depending
on the magnitude of the observed stars, and an accuracy of $\sim
1~\mu\mbox{as}$ is expected to be achieved in the more distant future. There
are two different ways in which astrometric microlensing signals can be used
to infer information: one possibility is to perform astrometric follow-up
observations on photometrically detected microlensing events, and the other
is to perform a survey based on astrometric observations alone. After the
predictable effects of the Sun and the planets, stars in the Galactic disk
play the dominant role in astrometric microlensing. The probability that the
disk stars introduce a centroid shift larger than the threshold $\delta_{\rm
T}$ at a given time for a given source in the Galactic bulge towards Baade's
window reaches 100\% for a threshold of $\delta_{\rm T} =
0.7~\mbox{$\mu$as}$, while this probability is
$\sim 2\%$ for 
$\delta_{\rm T} = 5~\mbox{$\mu$as}$.
However, this centroid shift does not {\em vary} much during the time in
which a typical photometric microlensing event differs from baseline. So
astrometric follow-ups (e.g.\ with SIM) are not expected to be disturbed by
the statistical astrometric microlensing due to disk stars, so that it is
possible to infer additional information about the nature of the lens that
caused the photometric event, as suggested. The probability to observe
astrometric microlensing events within the Galaxy turns out to be large
compared to photometric microlensing events. The probability to see a
variation by more than
$5~\mbox{$\mu$as}$ within one year and to reach the closest angular approach between lens
and source is $\sim 10^{-4}$ for a bulge star towards Baade's window,
while this reduces to $\sim 6\cdot 10^{-6}$ for a direction perpendicular to the Galactic plane.
For the upcoming mission GAIA, we expect
$\sim 1000$ of the observed stars to show a detectable astrometric microlensing
signal within its 5 year lifetime.  These events can be used to determine
accurate masses of the lenses, and to derive the mass and the scale
parameters (length and height) of the Galactic disk.

\end{abstract}

\keywords{astrometry -- galaxy: structure -- galaxy: stellar content --
gravitational lensing}

\section{Introduction}

It is known for more than one decade (Paczy{\'n}ski~\cite{Pac1}) that the
nature of matter between the observer and observed source stars can be
studied by observing brightenings of a large number of these stars caused by
the deflection of light by the intervening material. In addition to this
magnification effect, there is also a shift in the light centroid of the
observed star introduced by the lens object (H{\o}g, Novikov, \&
Polnarev~\cite{Hog}; Miyamoto
\& Yoshii~\cite{miyamoto}; Walker~\cite{walker}). 
Upcoming space missions will enable us to observe this centroid shift
(Paczy{\'n}ski~\cite{Pac2}; Boden, Shao, \& Van Buren~\cite{BSV}). In
particular, the Space Interferometry Mission (SIM, Allen et
al.~\cite{Allen1})\footnote{for information
about SIM see also {\tt
http://sim.jpl.nasa.gov}} will allow observations of selected targets with a
positional accuracy of $\sim 4~\mbox{$\mu$as}$ for sources brighter than $V
= 20$. Moreover, the Global Astrometric Interferometer for Astrophysics
mission (GAIA, Lindegren \& Perryman~\cite{lind})\footnote{for information
about GAIA see also {\tt
http://astro.estec.esa.nl/SA-general/Projects/GAIA/gaia.html}} will perform
an astrometric survey aimed at all-sky coverage (Gilmore et al. 1998) with
an accuracy of $20~\mbox{$\mu$as}$ ($60~\mbox{$\mu$as}$) for sources with
$V < 12$ ($V < 15$).\footnote{
Throughout the paper, we are talking about the accuracy of single astrometric 
measurements, not the accuracy of
parallax measurements obtained from the mission within its lifetime.} 
These two missions are somewhat complementary: While SIM has the ability to
point the instrument to selected targets, it will not perform a large survey program; on the other
hand, GAIA will perform an all-sky survey, but will not have the ability to point the instrument
to a selected target.

It has been mentioned (Paczy{\'n}ski~\cite{Pac2}; Boden et al.~\cite{BSV};
H{\o}g et al.~\cite{Hog}; Miyamoto
\& Yoshii~\cite{miyamoto}; Walker~\cite{walker}) that the observation
of the centroid shift during a (photometrically discovered)
microlensing event will yield additional information
about the lens, so that its mass, distance, and velocity can be determined
unambiguously.

Most of the discussions in the literature so far have been confined to the
centroid shifts of photometrically detected microlensing events which can be
detected by an instrument like SIM (e.g.\ Paczy{\'n}ski~\cite{Pac2}; Boden
et al.~\cite{BSV}).  It has been pointed out, however, that the microlensing
cross-section for centroid shift measurements is much larger than the
cross-section for light amplification (Paczy{\'n}ski~\cite{Pac3};
Miralda-Escud\'e~\cite{miralda}).  In this paper, we investigate the effects
of disk stars on the astrometric microlensing signal (centroid shift). The
disk stars can affect this signal in two ways.  First, for a microlensing
event that has been detected by its photometric signal, the intervening
matter can lead to additional centroid shifts and variations of these shifts
with time, which disturb the signal of the centroid shift caused by the lens
responsible for the photometrically detected microlensing event.  Second,
the disk stars form a population producing microlensing events that can be
detected by their astrometric microlensing signal alone in an astrometric
survey such as GAIA.

This paper is organized as follows. We discuss the signals of photometric
and astrometric microlensing in Sect.~2.  In Sect.~3, the optical depths due
to photometric and astrometric microlensing and the differences are
discussed. The characteristics of astrometric microlensing events and the
prospects for disk stars as lenses are discussed in Sect.~4. In Sect.~5, we
show that by observing astrometric microlensing events
 towards several directions, one can measure
the scale parameters of the mass distribution of the Galactic disk.
In Sect.~6, the effect of a luminous lens is discussed, while the implications for
upcoming space missions are discussed in Sect.~7. Finally, in
Sect.~8, the results of the
previous sections are summarized.

\section{The signals of photometric and astrometric microlensing}

\begin{figure*}
\epsfig{file=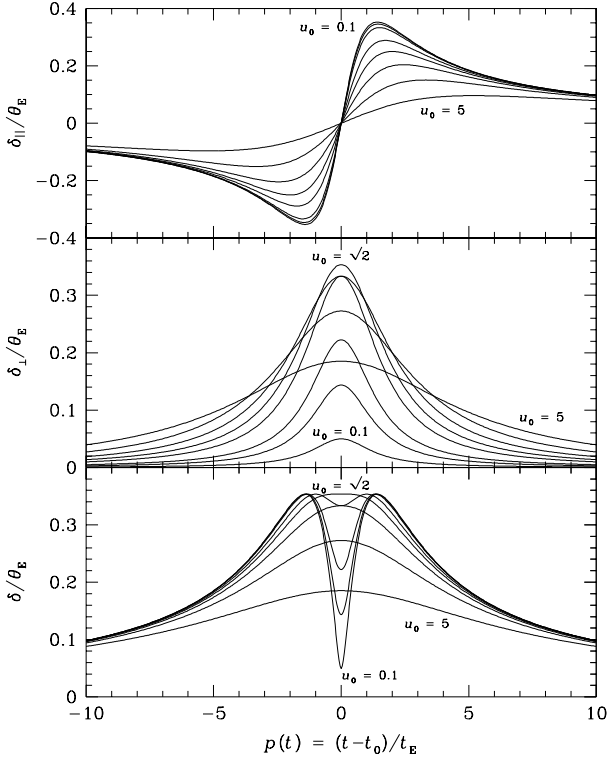,width=88mm}
\caption{The absolute centroid shift $\delta$ and its components along the direction of motion ($\delta_\|$)
and perpendicular to it ($\delta_\perp$) as a function of $p$ for the minimum separations
$u_0 = 0.1$, 0.3, 0.5, 1.0, $\sqrt{2}$, 2.0, 3.0, and 5.0.
{\em Top panel:} The parallel component $\delta_\|$. The curves are antisymmetric with respect to
$p=0$, the steepest curve corresponds to $u_0 = 0.1$ and the flattest curve to $u_0 = 5$.
{\em Middle panel:} The perpendicular component $\delta_\perp$. The curves are symmetric with respect to
$p=0$. At $p=0$, the largest value, namely $\sqrt{2}/4$, is reached for $u_0 = \sqrt{2}$; 
$u_0 = 1$ and $u_0 = 2$ yield the same $\delta_\perp(u_0,0)$.
For large given $|p|$, $\delta_\perp(u_0,p)$ decreases with smaller $u_0$.
{\em Bottom panel:} The absolute value of the centroid shift $\delta$. The curves are symmetric with
respect to $p=0$. The largest value at $p=0$ is reached for $u_0 = \sqrt{2}$, namely
$\sqrt{2}/4$. For $u_0 \geq \sqrt{2}$, there is a maximum at $p=0$, while for
$u_0 < \sqrt{2}$ a minimum occurs. $u_0 = 1$ and $u_0 = 2$ yield the same
$\delta(u_0,0)$.}
\label{deltafigs}
\end{figure*}

\begin{figure*}
\epsfig{file=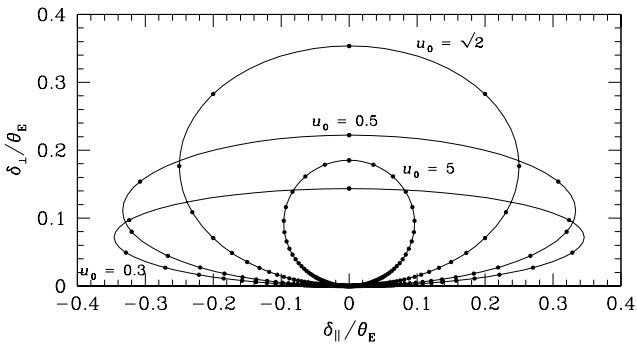,width=88mm}
\caption{The centroid shift trajectory in $(\delta_\|,\delta_\perp)$--space for
selected values of $u_0$ (where the brightness of the lens is neglected,
see text for details).
The centroid shift traces out an ellipse. The small dots on the trajectory mark
points of time spaced by $t_{\rm E}$.}
 \label{deltatraj}
\end{figure*}

Though the magnification of the source and the shift of its centroid of
light are based on the same effect, there are some qualitative differences
in the observable signals.

Let the source be located at a distance $D_{\rm{S}}$ from the observer and the lens with mass
$M$ at a distance
$0 < D_{\rm L} < D_{\rm S}$ from the observer.
Let $\vec \varphi_{\rm L}$ and $\vec \varphi_{\rm S}$ denote the angular positions of the lens and
source respectively. One can then define a dimensionless distance vector 
\begin{equation}
\vec u = \frac{\vec \varphi}{\theta_{\rm E}} =
\frac{\vec \varphi_{\rm S} - \vec \varphi_{\rm L}}{\theta_{\rm E}}\,,
\end{equation}
where
\begin{equation} 
\theta_{\rm E} = \sqrt{\frac{4GM}{c^2}\,\frac{D_{\rm S}-D_{\rm L}}{D_{\rm S}\,
D_{\rm L}}}
\end{equation}
is the angular Einstein radius. The Einstein radius
\begin{equation} 
r_{\rm E} = D_{\rm L}\,\theta_{\rm E} =
\sqrt{\frac{4GM}{c^2}\,\frac{D_{\rm L}\,(D_{\rm S}-D_{\rm L})}{D_{\rm S}}}
\end{equation}
gives the physical size of the angular Einstein radius in a plane perpendicular to the line-of-sight
observer-source at the position of the lens (lens plane).

In the following, we assume that there is no light contribution from an
unresolved luminous lens. The validity of this approximation and possible
modifications due to a luminous lens are discussed in
Sect.~\ref{blendsection}.

The magnification of the source due to the lens is given by
(e.g. Paczy{\'n}ski~\cite{Pac1})
\begin{equation}
\mu(u) = \frac{u^2 + 2}{u \sqrt{u^2+4}}\,,
\label{mueq}
\end{equation}
where $u = |\vec u|$.

For $u \ll 1$, one has
\begin{equation}
\mu(u) \simeq \frac{1}{u}\,,
\end{equation}
and for $u \gg 1$, one has
\begin{equation}
\mu(u) \simeq 1+\frac{2}{u^4}\,,
\end{equation}
so that for large angular separations, the lensed star produces a magnitude shift of
\begin{equation}
\Delta \mbox{mag} = -\frac{5}{\ln 10\,u^4}\,.
\label{magfalloff}
\end{equation}

The centroid shift of the source for a dark lens given by 
(H{\o}g et al.~\cite{Hog}; Miyamoto \& Yoshii~\cite{miyamoto}; Walker~\cite{walker}) 
\begin{equation}
\vec\delta(\vec u) = \frac{\vec u}{u^2+2}\,\theta_{\rm E}\,,
\label{deltaeq}
\end{equation}
i.e.\ it points away from the lens as seen from the source.

For $u \gg \sqrt{2}$, one has
\begin{equation}
\delta(u) \simeq \frac{1}{u}~\theta_{\rm E}\,,
\label{csfalloff}
\label{deltaapproxeq}
\end{equation}
so that the centroid shift falls off much more slowly than the magnitude shift towards larger $u$.
For $u \ll \sqrt{2}$,
\begin{equation}
\delta(u) \simeq \frac{u}{2}~\theta_{\rm E}\,,
\label{deltaapprox2eq}
\end{equation}
i.e. for small separations, the centroid 
shift tends linearly to zero, while the magnification increases
towards smaller separations. In contrast to the magnification, 
the absolute centroid shift reaches a maximum at $u = \sqrt{2}$ which is
\begin{equation}
\delta_{\rm max} =  \frac{\sqrt{2}}{4}\,\theta_{\rm E} \approx 0.354\,\theta_{\rm E}\,.
\end{equation}
While the magnification is a dimensionless scalar, the centroid shift is a vector with dimension, and
therefore it depends not only on the dimensionless separation $u$, but is also proportional to the
characteristic angular scale $\theta_{\rm E}$.
 
If one neglects the parallactic motion,
the relative path between lens and source is a straight line, so that 
\begin{equation}
u(t) = \sqrt{u_0^2 + [p(t)]^2}\,,
\end{equation}
where
\begin{equation}
p(t) = \frac{t-t_0}{t_{\rm E}}\,.
\label{eq:defp}
\end{equation}
This means that the closest approach between lens and source occurs at time $t_0$, where 
$|\vec u| = u_0$, and 
\begin{equation}
t_{\rm E} = \frac{\theta_{\rm E}}{\mu}\,,
\end{equation}
where $\mu$ is the relative proper motion between source and lens.

The absolute value of the centroid shift then reads
\begin{equation}
\delta(u_0,p) = \frac{\sqrt{u_0^2+p^2}}{u_0^2+p^2+2}\,\theta_{\rm E}\,,
\end{equation}
and the components against the direction of the motion of the lens relative to the source
$\delta_\|$ (i.e. in the direction of the motion of the source relative to the lens)
and perpendicular to it towards the side where the source is passed 
as seen by a moving
lens (i.e. away from the lens
as seen by a moving source)
$\delta_\perp$ are
\begin{eqnarray}
\delta_\|(u_0,p) & = & \frac{p}{u_0^2+p^2+2}\,\theta_{\rm E}\,,\nonumber\\
\delta_\perp(u_0,p) & = &\frac{u_0}{u_0^2+p^2+2}\,\theta_{\rm E}\,.
\end{eqnarray}

These functions are shown in Fig.~\ref{deltafigs} for several values of $u_0$.
While $\delta_\perp$ is symmetric around $p = 0$ and always positive, $\delta_\|$ is antisymmetric.
$\delta_\|$ has a maximum at $p=p_{{\rm m},\perp}$ and
a minimum at $p=-p_{{\rm m},\perp}$, where
\begin{equation}
p_{{\rm m},\perp} = \sqrt{u_0^2+2}
\end{equation}
and 
\begin{equation}
\delta_\|(u_0,p_{{\rm m},\perp}) = \frac{1}{2\,\sqrt{u_0^2+2}}\,\theta_{\rm E}\,.
\end{equation}
For $u_0 \ll 1$, one obtains
\begin{equation}
p_{{\rm m},\perp} \simeq \sqrt{2}
\end{equation}
and
\begin{equation}
\delta_\|(u_0,p_{{\rm m},\perp}) \simeq \delta_{\rm max}\,.
\end{equation}

$\delta_\perp$ reaches a maximum at $p=0$, 
where the height of the peak is maximal for $u_0 = \sqrt{2}$, reaching
$\delta_\perp = \delta_{\rm max}$, and in general the peak height is
\begin{equation}
\delta_\perp(u_0,0) = \frac{u_0}{u_0^2+2}\,\theta_{\rm E}\,.
\end{equation}

Since the absolute centroid shift has a maximum at $u = \sqrt{2}$, $\delta(u_0,p)$ goes through a minimum
at $p=0$ for  $u_0 < \sqrt{2}$ and has two maxima at $p = \pm p_{\rm m}$, where
\begin{equation} 
p_{\rm m} = \sqrt{2 - u_0^2}\,,
\end{equation}
i.e. $u = \sqrt{u_0^2+p^2} = \sqrt{2}$, so that
$\delta(u_0,p_m) = \delta_{\rm max}$. Note that for $u_0 \ll 1$, $p_{\rm m} \simeq \sqrt{2}$, so that
$p_{\rm m} \simeq p_{{\rm m},\perp}$.
For $u_0 \geq \sqrt{2}$, $\delta$ has only one maximum at $p=0$, where
\begin{equation}
\delta(u_0,0) = \delta_\perp(u_0,0) = \frac{u_0}{u_0^2+2}\,\theta_{\rm E}\,.
\end{equation}
Note that $\delta(u_1,0) = \delta(u_2,0)$ for $u_1 u_2 = 2$.

For large $|p|$, $\delta_\perp \propto 1/p^2$, while $\delta_\| \propto 1/p$, so that $\vec \delta$ points nearly
against the direction of the motion of the lens relative to the source for large $p$ 
and into it for small $p$, so that
the direction of the motion can be identified easily:
the change of the centroid shift is in the direction of the lens motion for
large $|p|$.
Due to the 
symmetry of $\delta_\perp$ and the antisymmetry of $\delta_\|$, the vector
\begin{equation}
\vec \delta_\perp = \frac{1}{2}\left(\vec \delta(u_0,p) + \vec \delta(u_0,-p)\right)
\end{equation}
points perpendicular to the lens motion relative to the source
towards the side where the source is passed and the vector 
\begin{equation}
\vec \delta_\| = \frac{1}{2}\left(\vec \delta(u_0,p) - \vec \delta(u_0,-p)\right)
\end{equation}
points against the direction of the motion of the lens relative to the source for $p > 0$ and into it
for $p < 0$.

In $(\delta_\|,\delta_\perp)$--space, the centroid-shift trajectory is an ellipse 
(e.g. Walker~\cite{walker})  
with semi-major axis $a$ in the $\delta_\|$-direction and semi-minor axis $b$ in the 
$\delta_\perp$-direction centered at $(0,b)$, where
\begin{equation}
a = \frac{1}{2}\,\frac{1}{\sqrt{u_0^2+2}}\,\theta_{\rm E}\,,\quad
b = \frac{1}{2}\,\frac{u_0}{u_0^2+2}\,\theta_{\rm E}\,.
\end{equation}
For $u_0 \to \infty$, this ellipse becomes a circle with radius $\theta_{\rm E}/(2u_0)$, and for
$u_0 \to 0$, the ellipse degenerates into a straight line of length $\theta_{\rm E}/\sqrt{2}$
(e.g. Walker~\cite{walker}). For selected values of $u_0$, the centroid-shift 
trajectory is shown in Fig.~\ref{deltatraj}.

\section{Optical depths for photometric and astrometric microlensing}
\label{sec:optdepth}

\begin{figure*}
\epsfig{file=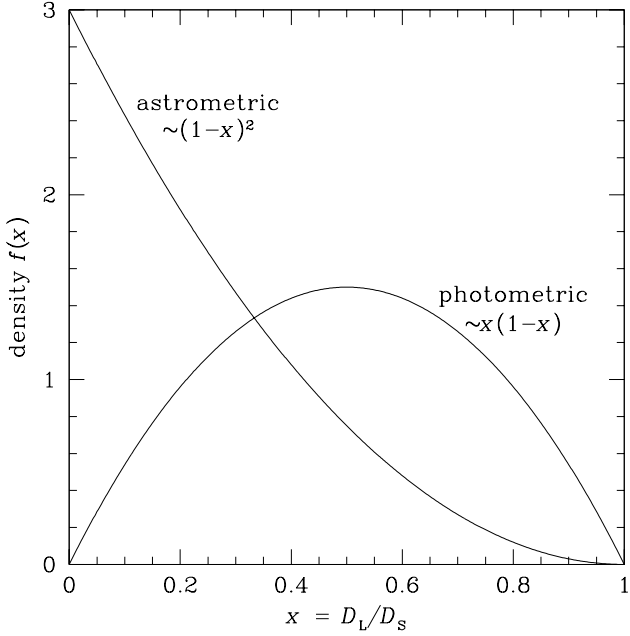,width=88mm}
\caption{Density functions $f(x)$ for photometric microlensing 
($f_1(x) = 6x(1-x)$) and astrometric
microlensing ($f_2(x) = 3(1-x)^2$) showing the favored and disfavored values for the lens distance
$D_{\rm L} = x\,D_{\rm S}$. The functions $f_i(x)$ are normalized, so that $\int_0^1 f_i(x) = 1$.}
\label{favfig}
\end{figure*}

Let $\sigma$ denote an area in the lens plane for which source positions
projected onto the lens plane yield a certain microlensing signature. The
probability $\gamma$ to observe such a signature for a given source is then
given by the product of the number area density of lenses and the area
$\sigma$. With $\rho(x)$ being the mass volume density at the distance
$D_{\rm L} = x\,D_{\rm S}$, and
$f_M(M)\,{\rm d}M$ being the distribution of lens masses, 
one obtains $\gamma$ due to lenses at any distance between source and observer as 
\begin{equation}
\gamma = D_{\rm S}\,\int\limits_0^1 \int\limits_0^{\infty} \frac{\rho(x)}{M}\,\sigma(x,M)\,
f_M(M)\,{\rm d}M\,{\rm d}x\,.
\label{probgen}
\end{equation}

For photometric microlensing, a commonly used signature is the magnification
of the light of the source star by more than a threshold $\mu_{\rm T}$ at a
given time, and the associated probability is referred to as {\rm optical
depth} of photometric microlensing $\tau_\mu$. This signature holds if the
angular separation between the lens and the source is smaller than a
corresponding threshold $u_{\rm T}$, given by the inversion of
Eq.~(\ref{mueq}) as
\begin{equation}
u_{\rm T} = \sqrt{\frac{2}{\sqrt{1-\mu_{\rm T}^{-2}}}-2}\,,
\end{equation}
and $u_{\rm T} = 1$ corresponds to $\mu_{\rm T} = 3/\sqrt{5} \approx 1.34$.
Therefore, $\sigma = \pi u_{\rm T}^2 r_{\rm E}^2$ in this case, and
the optical depth reads
\begin{equation}
\tau_\mu(u_{\rm T}) = u_{\rm T}^2\,\tau_\mu(1)\,,
\end{equation}
where
\begin{equation}
\tau_{\mu}(1) = \frac{4\pi G}{c^2} D_{\rm S}^2 \int\limits_0^1 \rho(x)\,x(1-x)\,{\rm d}x\,.
\end{equation}
Note that $\tau_\mu$ does not depend on the masses of the lenses and that,
in addition to distances with large $\rho$, objects around half-way between
the observer and the source are favored.

Let us now consider a similar signature for the centroid shift, namely 
the case where the centroid shift exceeds a given threshold
$\delta_{\rm T}$.
From Eq.~(\ref{deltaeq}) one obtains that the absolute centroid shift exceeds 
a given threshold
$\delta_{\rm T}$ if $u \in [u_{\rm T}^{-}, u_{\rm T}^{+}]$, where 
\begin{equation}
u_{\rm T}^{\pm} = \frac{\theta_{\rm E}}{2\delta_{\rm T}} \pm
\sqrt{\frac{\theta_{\rm E}^2}{4 \delta_{\rm T}^2}-2}
\end{equation}
and $u_{\rm T}^{+} > \sqrt{2} > u_{\rm T}^{-}$ 
for $\delta_{\rm T} < \delta_{\rm max} = (\sqrt{2}/4)\,\theta_{\rm E}$. Otherwise, there are no solutions due to
the fact that $\delta_{\rm T}$ cannot exceed 
$\delta_{\rm max}$.
Since the centroid shift is not a dimensionless quantity, $u_{\rm T}^{\pm}$
depend on 
$\theta_{\rm E}$, whereas for
photometric signatures, $u_{\rm T}$
depends only on the magnification threshold $\mu_{\rm T}$ and not on any other quantity.
For $K = \theta_{\rm E}/\delta_{\rm T} \gg 1$
\begin{equation}
u_{\rm T}^{+} \simeq \frac{\theta_{\rm E}}{\delta_{\rm T}}\,, \quad
u_{\rm T}^{-} \simeq 2 \frac{\delta_{\rm T}}{\theta_{\rm E}}\,, 
\end{equation}
which also correspond to the large separation and small separation limits, 
Eqs.~(\ref{deltaapproxeq})
and~(\ref{deltaapprox2eq}).
As we will see in more detail in the next section, $\theta_{\rm E}$ is of order
$\mbox{mas}$, while $\delta_{\rm T}$ is of order $\mu\mbox{as}$, so that this is a fair approximation.

Since the area in the lens plane giving a centroid shift larger than $\delta_{\rm T}$ is given
by
$\pi [(u_{\rm T}^{+})^2 - (u_{\rm T}^{-})^2]\,r_{\rm E}^2$, and 
\begin{equation}
(u_{\rm T}^{+})^2 - (u_{\rm T}^{-})^2
= K^2\,\sqrt{1-\frac{8}{K^2}} \simeq 
K^2 -4 - \frac{8}{K^2}\,, 
\end{equation}
the optical depth  for centroid shifts larger than $\delta_{\rm T}$ can be written as
\begin{equation}
\tau_\delta(u_{\rm T}^{-},u_{\rm T}^{+}) \simeq \tau_\delta
(0,\theta_{\rm E}/\delta_{\rm T}) -4 \tau_\mu(1)\,,
\end{equation}
i.e. the corresponding area can be approximated by a circle with radius 
$u_{\rm T} = \theta_{\rm E}/\delta_{\rm T}$, so that the upper threshold becomes $u_{\rm T}^{+} 
	\simeq u_{\rm T}$ and the lower threshold becomes $u_{\rm T}^{-} \simeq 0$.
This means that $\sigma = \pi\,u_{\rm T}^2\,r_{\rm E}^2 = \pi\,D_{\rm L}^2\,\theta_{\rm E}^4/\delta_{\rm T}^2$ and 
with Eq.~(\ref{probgen}), $\tau_\delta$ reads
\begin{eqnarray}
\tau_\delta(0,\theta_{\rm E}/\delta_{\rm T}) & = &
\pi D_{\rm S} \int\limits_0^1 \int\limits_0^{\infty} \frac{\rho(x)}{M}\,
\frac{D_{\rm L}^2\,\theta_{\rm E}^4}{\delta_{\rm T}^2}\,
f_M(M)\,{\rm d}M\,{\rm d}x\,
 \nonumber \\
& = & \frac{16\pi G^2}{c^4}\, \frac{D_{\rm S}\,\overline{M}}{\delta_{\rm T}^2}\,
\int\limits_0^{1} \rho(x)\,(1-x)^2\,{\rm d}x\,,
\label{taudeltaeq}
\end{eqnarray}
where 
\begin{equation}
\overline{M} = \int\limits_0^{\infty} M\,f_M(M)\,{\rm d}M 
\end{equation}
is the average mass from the mass spectrum $f_M$.

Contrary to photometric microlensing, small distances between observer and
lens are favored, so that disk stars give the main contribution. In
addition, large distances between observer and lens are disfavored compared
to photometric microlensing (see Fig.~\ref{favfig}). While the bulge stars
and the LMC stars may play an important role in the photometric microlensing
towards the bulge (Kiraga \& Paczy{\'n}ski~\cite{kiraga}) and the LMC
(Sahu~\cite{sahu}), respectively, their contribution to astrometric
microlensing is very small.

From the expression for the optical depth $\tau_\delta$, Eq.~(\ref{taudeltaeq}), one sees 
that a probability
density for a lens yielding  a deflection above a given threshold at any time is given by
\begin{equation}
f_x(x) = C_0\,\frac{{\rm d}\tau_\delta}{{\rm d}x} = C_1\,\rho(x)\,(1-x)^2\,,
\label{lensdisteq}
\end{equation}
so that the expectation value for the lens distance is given by
\begin{equation}
<\!\!x\!\!> = \frac{\int_0^1 \rho(x)\,x(1-x)^2{\rm d}x}{\int_0^1 \rho(x)\,(1-x)^2\,{\rm d}x}\,.
\end{equation}
For constant mass density $\rho(x) = \rho_0$, one obtains
\begin{equation}
<\!\!x\!\!> = \frac{\int_0^1 x(1-x)^2{\rm d}x}{\int_0^1 (1-x)^2\,{\rm d}x} = \frac{1}{4}\,.
\end{equation}

After having established that the main contribution comes from close lenses, 
we can estimate the detection threshold for $D_{\rm S} \gg D_{\rm L}$: The
angular Einstein radius $\theta_{\rm E}$ reads in this limit
\begin{eqnarray}
\theta_{\rm E} & = & \sqrt{\frac{4GM}{c^2\,D_{\rm L}}} \nonumber \\
& = & 
2.0
\left(\frac{M}{0.5~M{\sun}}\right)^{1/2}\,\left(\frac{D_{\rm L}}{1~\mbox{kpc}}\right)^{-1/2}\,
\mbox{mas}\,,
\label{thetaEphys}
\end{eqnarray}
For the maximum separation $u_{\rm T}$ yielding a signal above the threshold
$\delta_T$, one obtains
\begin{equation}
u_{\rm T} \simeq 2000\,
\left(\frac{M}{0.5~M_{\sun}}\right)^{1/2}\,
\left(\frac{D_{\rm L}}{1~\mbox{kpc}}\right)^{-1/2}\,
\left(\frac{\delta_{\rm T}}{1~\mu\mbox{as}}\right)^{-1}\,.
\label{utexp}
\end{equation}
Note that this is a gigantic number compared to
photometric microlensing which yields a magnification 1\% above the
baseline for $u = 3.8$, while for $u=200$, the
magnification is only by a factor $1.4\cdot 10^{-9}$ above the baseline,
and for $u=2000$, this reduces to $1.4\cdot 10^{-13}$.

\begin{table*}
\begin{center}
\caption[]{Astrometric microlensing optical depth} 
\label{table:optdepth}
\begin{tabular}{ccc}
\tableline
 \tableline
 & \multicolumn{2}{c}{Astrometric microlensing optical depth}
 \\
Detection & \multicolumn{2}{c}{per observed star}
 \\
threshold &\multicolumn{2}{c}{-----------------------------------------------------------------------------------------} \\
&  
Bulge stars towards Baade's window & Perpendicular to Galactic plane \\
& $\rho(x) = \rho_0$ & $\rho(x) = \rho_0\,\exp\{-x\,D_{\rm S}/H\}$  \\
& $D_{\rm S} = 8.5~\mbox{kpc}$ & $D_{\rm S} \gg H = 300~\mbox{pc}$ \\
& & \\
$\delta_{\rm T}$ ($\mu$as) & 
 $\tau_{\delta,0}$&
$\tau_{\delta,\infty}$ \\
\tableline
$0.7$ & $1.0$ & $0.11$\\
$1$   & $0.55$ &$5.8\cdot 10^{-2}$  \\
$5$   & $2.2\cdot10^{-2}$& $2.3\cdot 10^{-3}$ \\
$10$  & $5.5\cdot10^{-3}$& $5.8\cdot 10^{-4}$ \\ 
$100$  & $5.5\cdot10^{-5}$ & $5.8\cdot 10^{-6}$ \\
\tableline
\end{tabular}
\tablecomments{The astrometric microlensing optical depth 
$\tau_\delta \propto \rho_0\,\overline{M}\,\delta_{\rm T}^{-2}$
is shown as a function of the detection threshold $\delta_{\rm T}$  
for sources (i) towards the Galactic bulge and (ii) perpendicular
to the Galactic plane, with the reference values 
$\overline{M} = 0.5~M_{\sun}$ and $\rho_0 = 0.08~M_{\sun}\,\mbox{pc}^{-3}$,
Eqs.~(\ref{taudeltavals}) and~(\ref{taudeltavalsexp}).}
\end{center}
\end{table*}

Let us also look how the centroid shift varies with time, i.e.\
consider a variation in the angular separation between lens and source described by 
a proper motion $\mu = {\rm d}\varphi/{\rm d}t = v/D_{\rm L}$.
Assuming the lens to be dark or resolved from the source, the change in the 
centroid shift is given by
\begin{equation}
\frac{{\rm d}\delta}{{\rm d}u} = \frac{2-u^2}{\left(u^2+2\right)^2}\,\theta_{\rm E}\,,
\end{equation}
which gives for $u \gg 1$ 
\begin{equation}
\frac{{\rm d}\delta}{{\rm d}u} \simeq -\frac{1}{u^2}\,\theta_{\rm E}\,,
\end{equation}
or expressed with $\varphi = u\,\theta_{\rm E}$
\begin{equation}
\frac{{\rm d}\delta}{{\rm d}\varphi} \simeq  
- \frac{\theta_{\rm E}^2}{\varphi^2} = -\frac{1}{u^2}\,,
\end{equation}
i.e. the change of the centroid shift with the distance falls off one power 
faster than the centroid shift itself, Eq.~(\ref{csfalloff}),
however, 2 powers slowlier than the shift in magnitude, Eq.~(\ref{magfalloff}). With
\begin{equation}
\frac{{\rm d}\varphi}{{\rm d}t} = 58\,
\left(\frac{v}{100~\mbox{km~s}^{-1}}\right)\,\left(\frac{D_{\rm L}}{1~\mbox{kpc}}\right)^{-1}\,
\mbox{$\mu$as~days}^{-1}\,, 
\end{equation}
one gets
\begin{eqnarray}
&&\frac{{\rm d}\delta}{{\rm d}t}  = 
\frac{{\rm d}\delta}{{\rm d}\varphi}\, 
\frac{{\rm d}\varphi}{{\rm d}t} \nonumber \\ && \!\!\!\!\!\!\!\!\!\!\!\!\!
=  - 58\,\frac{\theta_{\rm E}^2}{\varphi^2}\, 
\left(\frac{v}{100~\mbox{km~s}^{-1}}\right)\,\left(\frac{D_{\rm L}}{1~\mbox{kpc}}\right)^{-1}\,
\mbox{$\mu$as~days}^{-1}\,. 
\end{eqnarray}
For $D_{\rm S} \gg D_{\rm L}$, the
angular Einstein radius $\theta_{\rm E}$ is given by Eq.~(\ref{thetaEphys})
and the time in which the angular separation between lens and source changes by 
$\theta_{\rm E}$ is
given by
\begin{eqnarray}
t_{\rm E}  & = & \theta_{\rm E}/\mu \nonumber \\
 & = &
35
\left(\frac{M}{0.5~M{\sun}}\right)^{1/2}\,\left(\frac{D_{\rm L}}{1~\mbox{kpc}}\right)^{1/2}\,
\times \nonumber \\
& & \times\,
\left(\frac{v}{100~\mbox{km~s}^{-1}}\right)^{-1}\,
\label{tEphys}
\mbox{days}\,. 
\end{eqnarray}
This means that for a close encounter at 
a minimal angular separation of $\lesssim 1~\theta_{\rm E}$, one has
still a centroid shift of $\sim 2~\mbox{$\mu$as}$ at a time $t = 1000~t_{\rm E} \sim 100~\mbox{yr}$ after the
closest encounter, a centroid shift of
$\sim 20~\mbox{$\mu$as}$ at a time $t = 100~t_{\rm E} \sim 10~\mbox{yr}$,
and  a centroid shift of
$\sim 200~\mbox{$\mu$as}$ at a time $t = 10~t_{\rm E} \sim 1~\mbox{yr}$, where the magnitude shift
is only of the order of $10^{-4}$.

Since large contributions to the optical depth of astrometric microlensing
are expected for small distances between lens and observer, the disk stars
are expected to play the most important role regardless of where the source
star is located.\footnote{In fact, an even larger role is played by the sun
and the solar planets, whose effect is being taken into account in the SIM
mission (R.~J.~Allen 1998, private communication).}

For sources in the Galactic bulge towards Baade's window ($l = -1^{\circ}$,
$b = -4^{\circ}$), which is the field of interest for the photometric
microlensing surveys towards the Galactic bulge, the mass density of the
disk stars is approximately constant, so that the optical depth for centroid
shifts larger than $\delta_{\rm T}$ reads
\begin{eqnarray}
\tau_{\delta,0} 
& = & \frac{16\pi G^2}{c^4}\, D_{\rm S}\,\frac{\overline{M}\,\rho_0}{\delta_{\rm T}^2}\,
\int\limits_0^{1} (1-x)^2\,{\rm d}x \nonumber \\
& = & \frac{16\pi G^2}{3\,c^4}\, D_{\rm S}\,\frac{\overline{M}\,\rho_0}{\delta_{\rm T}^2} \nonumber \\
& = & 0.55\,\left(\frac{D_{\rm S}}{8.5~\mbox{kpc}}\right)\,\left(\frac{\overline{M}}{0.5~M_{\sun}}\right)\,
\times \nonumber \\
& & \times\,
\left(\frac{\rho_0}{0.08~M_{\sun}\,\mbox{pc}^{-3}}\right)
\,\left(\frac{\delta_{\rm T}}{1~\mbox{$\mu$as}}\right)^{-2}\,.
\label{taudeltavals}
\end{eqnarray}

Values for $\tau_{\delta,0}$ for the reference values of $\overline{M}$ and
$\rho_0$ 
are shown in Table~\ref{table:optdepth}
for several values of $\delta_{\rm T}$.
If one compares these values to the optical depth for light amplification
\begin{equation}
\tau_\mu(1) = 5.8\cdot 10^{-7}\,
\left(\frac{D_{\rm S}}{8.5~\mbox{kpc}}\right)^2\,
\left(\frac{\rho_0}{0.08~M_{\sun}\,\mbox{pc}^{-3}}\right)\,,
\end{equation}
one sees that $\tau_\delta \gg 4 \tau_\mu(1)$ and therefore
the approximation $\tau_\delta(u_{\rm T}^{-}, u_{\rm T}^{+}) 
\approx \tau_\delta(0,\theta_{\rm E}/\delta_{\rm T})$ is justified.

The case of an exponential behaviour of the mass density is discussed in Sect.~\ref{diskscales}.

\section{Astrometric microlensing events}

\begin{figure*} 
\epsfig{file=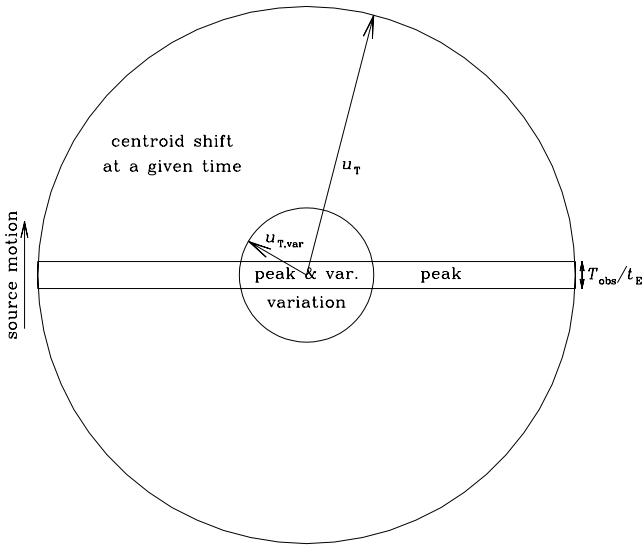,width=88mm}
\caption{Regions $\sigma$ in the lens plane that correspond to projected 
source positions that yield a given
signature. All distances are given in multiples of the Einstein radius $r_{\rm E}$. The lens is
located in the center of the figure. The source moves in the indicated direction during the
observation time $T_{\rm obs}$, and the regions $\sigma$ have been positioned with respect to the source
position at the midpoint between the beginning and the end of the observations.
The outer circle with radius $u_{\rm T} = \theta_{\rm E}/\delta_{\rm T}$ includes the
positions of the source where the centroid shift exceeds the threshold $\delta_{\rm T}$,
the inner circle with radius $u_{\rm T,var} =
[(\theta_{\rm E}\,T_{\rm obs})/(\delta_{\rm T}\,t_{\rm E})]^{1/2}$ includes the source positions for
which the variation of the centroid shift during $T_{\rm obs}$ exceeds
$\delta_{\rm T}$. Only for smaller regions, the closest approach between lens and
source occurs within $T_{\rm obs}$ yielding a peak signature.}
\label{scheme}
\end{figure*}

\subsection{The characteristics}
Photometric microlensing is described by 3 characteristic quantities: The optical depth $\tau$, 
the event rate $\Gamma$, 
and the average duration of an event $<\!\!t_{\rm e}\!\!>$, where one defines an event to last if the 
magnification exceeds a given threshold $\mu_{\rm T}$. These three characteristics are related by
(Griest~\cite{Griest})
\begin{equation}
\tau = \Gamma<\!\!t_{\rm e}\!\!>\,.
\end{equation}
Consider coordinates in the lens plane, where the lens is at rest and the projected position
of the source moves with a velocity $v = D_{\rm L}\,\mu$.
As discussed in Sect.~\ref{sec:optdepth}, the magnification exceeds $\mu_{\rm T}$, if the position
of the source projected onto the lens plane is in a circle of radius $u_{\rm T}\,r_{\rm E}$ 
around the lens.
Optical depth, event rate, and average event duration can be related to the 
`area', `width', and 
`average length' of this circle, respectively (Mao \& Paczy{\'n}ski~\cite{MP91}; Dominik~\cite{DoThesis}).
The area is given by $a = \pi\,u_{\rm T}^2\,r_{\rm E}^2$.
The width $w$ is given by the range of impact parameters for which a moving
source hits the area, in this case $w = 2\,u_{\rm T}\,r_{\rm E}$. The average length $\overline{l}$ is
given by the average length of the portion of the source trajectory where the source is inside the
area, in this case $\overline{l} = \frac{\pi}{2}\,u_{\rm T}\,r_{\rm E}$.

For the optical depth, the area of successful source positions is given by $\sigma_\tau =
a = \pi\,u_{\rm T}^2\,r_{\rm E}^2$. 
All sources within a rectangle with sides $w = 2\,u_{\rm T}\,r_{\rm E}$ (perpendicular to
the motion) and
$T_{\rm obs}\,v$ (parallel to the motion) 
will reach their closest approach to the lens within $T_{\rm obs}$ and thereby 
show a peak in their light curve. The area corresponding to events that peak
within $T_{\rm obs}$ is therefore
$\sigma_{\rm peak} = 2\,u_{\rm T}\,r_{\rm E}\,T_{\rm obs}\,v$. Since every source that enters the
area given by $\sigma_\tau$ peaks exactly once, the event rate is given by
$\Gamma = \gamma_{\rm peak}/T_{\rm obs}$.
The average event duration is finally given by $<\!\!t_{\rm e}\!\!> = \overline{l}/v = \frac{\pi}{2}\,u_{\rm T}\,t_{\rm E}$.

For photometric microlensing, $u_{\rm T} \sim 1$, and $<\!\!t_{\rm e}\!\!> 
\sim t_{\rm E} \sim 1~\mbox{month}$, i.e.\ for
$T_{\rm obs} \sim 1~\mbox{yr}$, $<\!\!t_{\rm e}\!\!>\, \ll T_{\rm obs}$.
This means that one observes the events from baseline to peak and back to baseline. This implies that an event with
a peak amplification of $A_{\rm peak}$ brightens by this amplification and fades back within $T_{\rm obs}$, i.e.\ events
that reach $A_{\rm T}$ also vary by $A_{\rm T}$ or more within the observation time.

Since $u_{\rm T} \gg 1$ for typical astrometric events, the situation is quite different. Though in both cases, photometric
and astrometric microlensing, only the variation of the signal (the magnification or the centroid shift) can be
observed, not the signal itself, this difference strongly affects astrometric microlensing, while it does not affect
photometric microlensing, unless $t_{\rm E}$ is very long. 
For astrometric microlensing, 
$<\!\!t_{\rm e}\!\!> \sim 200\,t_{\rm E} \sim 20~\mbox{yr}$, which 
may well exceed $T_{\rm obs}$, so that one has to look for
configurations where the signal {\em varies} by a given amount rather than for configurations where it exceeds some
amount (compared to an asymptotic value which is unknown in this case).

As shown later, for small $\delta_{\rm T}$ ($\lesssim 10~\mbox{$\mu$as}$) and 
$T_{\rm obs} \lesssim 10~\mbox{yr}$, the region of source positions for which the centroid shift
varies by more than $\delta_{\rm T}$ within $T_{\rm obs}$ can be approximately described
by a circle of radius $u_{\rm T,var}\,r_{\rm E}$, where $u_{\rm T,var} < u_{\rm T}$, and
$u_{\rm T,var} \to u_{\rm T}$ for $T_{\rm obs} \to \infty$.
Therefore, one has an analogous situation to the case where the criterion that the
centroid shift exceeds $\delta_{\rm T}$ is used: $u_{\rm T}$ just needs to be replaced by
$u_{\rm T,var}$. While the region $\sigma_{\rm var} = \pi\,u_{\rm T,var}^2\,r_{\rm E}^2$ 
corresponds to source positions giving rise to centroid shift variations larger
than $\delta_{\rm T}$ within
$T_{\rm obs}$, this does not give the event rate, because the same event may show a significant
variation within subsequent time intervals. Instead, it is again useful to consider the
closest approach between lens and source to occur within $T_{\rm obs}$ yielding a peak
signature. The source positions yielding a significant variation and a peak signature
are located within a rectangle with sides $u_{\rm T}\,r_{\rm E}$ (perpendicular to motion) and
$T_{\rm obs}\,v$ (parallel to motion), so that the area of successful source positions is
$\gamma_{\rm var,peak} = 2\,u_{\rm T,var}\,r_{\rm E}\,T_{\rm obs}\,v$. Figure~\ref{scheme}
illustrates
the regions yielding the different signatures.

While for $<\!\!t_{\rm e}\!\!> \ll T_{\rm obs}$, the observed variation
becomes identical with the maximum signal, for 
$<\!\!t_{\rm e}\!\!> \gtrsim T_{\rm obs}$
it can happen that one sees a significant variation
without reaching the peak and that one reaches the peak but does not see a significant variation.
For the actual centroid shift
being much larger than $\delta_{\rm T}$ and the closest approach
being reached within $T_{\rm obs}$, 
the observed {\it variation} of the centroid shift during $T_{\rm obs}$ may fall below
the threshold.  
On the other hand,
the variation in the centroid shift can be larger than $\delta_{\rm T}$ 
without reaching the maximal value within $T_{\rm obs}$. In such a
case, a monotonous variation of the centroid shift can be
seen, which moreover points approximately into the same direction. 
The observed centroid of light also moves due to the proper motion of the
source (and a luminous lens) and due to the parallactic motion and these motions have to be corrected
for. In fact, for $u \gg 1$, the proper motion  can be many orders of 
magnitude larger than the centroid shift due to lensing. The centroid shift
due to lensing  can only be separated by detecting its different time behaviour.
Therefore, the subset of events that also `peak' within $T_{\rm obs}$ forms a class of events
with a signature that is distinct from proper motion (and parallactic motion) and
hence can be more easily detected and distinguished.

Note that the effective observation time can be substantially stretched just by making a few
additional measurements after a few years.

\subsection{Significant variation in an event}

\begin{table*}
\begin{center}
\caption[]{Probability of observing a significant centroid-shift variation}
\label{table:gamvar}
\begin{tabular}{ccc}
\tableline
\tableline
 & \multicolumn{2}{c}{Probability of observing a centroid-shift variation}
 \\
Detection & \multicolumn{2}{c}{larger than $\delta_{\rm T}$ 
within $T_{\rm obs} = 1~\mbox{yr}$ for a given observed star}
 \\
threshold &\multicolumn{2}{c}{-----------------------------------------------------------------------------------------} \\
&  
Bulge stars towards Baade's window & Perpendicular to Galactic plane \\
& $\rho(x) = \rho_0$ & $\rho(x) = \rho_0\,\exp\{-x\,D_{\rm S}/H\}$  \\
& $D_{\rm S} = 8.5~\mbox{kpc}$ & $D_{\rm S} \gg H = 300~\mbox{pc}$ \\
& & \\
$\delta_{\rm T}$ ($\mu$as) & 
$\gamma_{{\rm var},0}$& $\gamma_{{\rm var},\infty}$\\
\tableline
$1$ & $4.3\cdot 10^{-3}$&$3.0\cdot 10^{-4}$  \\
$5$ & $8.6\cdot 10^{-4}$&$6.0\cdot 10^{-5}$  \\
$10$ &$4.3\cdot 10^{-4}$&$3.0\cdot 10^{-5}$  \\
$100$ & $4.3\cdot 10^{-5}$&$3.0\cdot 10^{-6}$ \\ 
\tableline
\end{tabular}
\tablecomments{The probability of observing a variation in the
centroid shift 
larger than the threshold  
$\delta_{\rm T}$, 
$\gamma_{\rm var} \propto \rho_0\,T_{\rm obs}\,v\,\delta_{\rm T}^{-1}$, is shown  
for sources (i) towards the Galactic bulge, Eq.~(\ref{ratevar}), and (ii) perpendicular
to the Galactic plane, Eq.~(\ref{ratevarexp}), with the reference values $T_{\rm obs} = 1~\mbox{yr}$, $v = 100$ km s$^{-1}$,  
and $\rho_0 = 0.08~M_{\sun}\,\mbox{pc}^{-3}$.
}
\end{center}
\end{table*}

Let us investigate the change of centroid shift between two points of time separated by 
$T_{\rm obs}$. Let $p_0$ denote the value of $p(t)$ (c.f.\ Eq.~(\ref{eq:defp})) in the middle between these points and
\begin{equation}
\Delta p = \frac{T_{\rm obs}}{2\,t_{\rm E}}\,.
\end{equation}
For $u_0^2 + p_0^2 \gg 1$, the square of the absolute value of the change in centroid shift is given by
\begin{eqnarray}
& & \!\!\!\!\!\!\!\!\!\!\!\!D^2(u_0,p_0-\Delta p,p_0+\Delta p)\; = \nonumber \\ 
& = & \left| \vec \delta(u_0,p_0+\Delta p) - \vec \delta(u_0,p_0-\Delta p)\right|^2 \nonumber \\
 & = & \left[\left(\frac{u_0}{u_0^2+(p_0+\Delta p)^2} - \frac{u_0}{u_0^2+(p_0-\Delta p)^2}\right)^2
 +\right. \nonumber \\
& & \!\!\!\!\!\!\!\!\!\!\!\!\!\!
+\,\left. \left(\frac{p_0+\Delta p}{u_0^2+(p_0+\Delta p)^2} - \frac{p_0-\Delta p}{u_0^2+(p_0-\Delta p)^2}\right)^2
	\right]\,\theta_{\rm E}^2 \,.
\label{delcent}
\end{eqnarray}
For $(\Delta p)^2 \ll u_0^2+p_0^2$ one obtains
\begin{equation}
D^2 = 4 \frac{(\Delta p)^2\,\theta_{\rm E}^2}{\left(u_0^2+p_0^2\right)^2}\,.
\label{DD}
\end{equation}
In this limit, $D$ is also the maximum change in the centroid shift within $T_{\rm obs}$ around
$p_0$,\footnote{For $u_0 \gg 1$, the centroid-shift curve in space is a circle, 
so that the largest difference between
two points within the traced time is the difference of the centroid-shift vectors at the boundary points
if less than half the circumference is traced and the largest difference is equal to the diameter of the circle
if half of the circumference or more is traced. For small $\Delta p$, 
one traces less than half the circumference. For $p^2 \geq u_0^2+2$, both components of $\vec \delta$ fall
monotonously, so that the largest difference also occurs between the boundary points for small $u_0$ but
larger $|p|$.} so that the condition for a change in the centroid shift above the threshold means that
$(u_0,p_0)$ lie within a circle of radius
\begin{eqnarray}
u_{\rm T,var} & = &\sqrt{\frac{2\,\Delta p\,\theta_{\rm E}}{\delta_{\rm T}}} \nonumber \\
 & = & \sqrt{\frac{T_{\rm obs}\,\theta_{\rm E}}{\delta_{\rm T}\,t_{\rm E}}} \nonumber \\
& = & \sqrt{\frac{T_{\rm obs}\,v}{\delta_{\rm T}\,D_{\rm L}}}\,.
\label{varthresh}
\end{eqnarray}
Using reference values, $u_{\rm T,var}$ reads
\begin{eqnarray}
u_{\rm T,var} & = & 144\, 
\left(\frac{\delta_{\rm T}}{1~\mu\mbox{as}}\right)^{-1/2}\,
\left(\frac{D_{\rm L}}{1~\mbox{kpc}}\right)^{-1/2}\,
\times \nonumber \\
& & \times\,
\left(\frac{v}{100~\mbox{km~s}^{-1}}\right)^{1/2}\,
\left(\frac{T_{\rm obs}}{1~\mbox{yr}}\right)^{1/2}\,,
\label{utvarexp}
\end{eqnarray}
which can be much smaller than $u_{\rm T}$, though still $u_{\rm T,var} \gg 1$.
Let us now check the assumption $(\Delta p)^2 \ll u_{\rm T,var}^2$, which becomes
\begin{equation}
\frac{T_{\rm obs}\,\mu}{\theta_{\rm E}} \ll 4 \frac{\theta_{\rm E}}{\delta_{\rm T}}
\label{approx1}
\end{equation}
with Eq.~(\ref{varthresh}), i.e. the change in the angular separation between lens and source in units of angular Einstein radii 
is much smaller than the ratio between (4 times) the angular Einstein radius and the
centroid shift threshold $\delta_{\rm T}$.
Eq.~(\ref{approx1}) can also be written as
\begin{equation}
F = \frac{\delta_{\rm T}\,T_{\rm obs}\,v}{4\,\theta_{\rm E}^2\,D_{\rm L}} \ll 1\,.
\end{equation}
With $<\!\!t_{\rm e}\!\!> = \frac{\pi}{2}\,u_{\rm T}\,t_{\rm E}$, $u_{\rm T} = \theta_{\rm E}/
\delta_{\rm T}$, and $t_{\rm E} = (D_{\rm L}\,\theta_{\rm E})/v$, one sees that $F \ll 1$ reflects
the condition $<\!\!t_{\rm e}\!\!> \gg T_{\rm obs}$.
For nearby lenses ($D_{\rm S} \gg D_{\rm L}$), one obtains
\begin{eqnarray}
F & = &\frac{c^2\,\delta_{\rm T}\,T_{\rm obs}\,v}{16 GM} \\
& = & 1.3\cdot 10^{-3}\,
\left(\frac{M}{0.5~M_{\sun}}\right)^{-1}\,
\left(\frac{T_{\rm obs}}{1~\mbox{yr}}\right)\,
\times \nonumber \\
& & \times\,
\left(\frac{v}{100~\mbox{km s}^{-1}}\right)\,
\left(\frac{\delta_{\rm T}}{1~\mbox{$\mu$as}}\right)\,,
\end{eqnarray}
so that the condition $F \ll 1$ is fulfilled for
$\delta_{\rm T} \lesssim 10~\mbox{$\mu$as}$ and $T_{\rm obs} \lesssim 10~\mbox{yr}$.

The next-order corrections to the circle $u_0^2 + p_0^2 = u_{\rm T,var}^2$ can be determined
by looking at the cases $u_0 = 0$ and $p_0 = 0$.
For $u_0 = 0$, one obtains from Eq.~(\ref{delcent})
\begin{equation}
D^2 = \frac{4 (\Delta p)^2}{\left[p_0^2 - (\Delta p)^2\right]^2}\,\theta_{\rm E}^2\,,
\end{equation}
so that the threshold $\delta_{\rm T}$ is reached for ($p_0 > \Delta p$)
\begin{eqnarray}
p_{0,{\rm T}}^2 & = & 
\frac{T_{\rm obs}\,v}
{\delta_{\rm T}\,D_{\rm L}}\,
\left(1 + \frac{T_{\rm obs}\,v\,\delta_{\rm T}}{4\,\theta_{\rm E}^2\,D_{\rm L}}\right) 
\nonumber \\
& = & 
\frac{T_{\rm obs}\,v}{\delta_{\rm T}\,D_{\rm L}}\,
\left(1 + F\right)\,,
\end{eqnarray} 
which reveals Eq.~(\ref{varthresh}) for $F \ll 1$.

For $p_0 = 0$, one obtains
\begin{equation}
D^2 = \frac{4 (\Delta p)^2}{\left[u_0^2 + (\Delta p)^2\right]^2}\,\theta_{\rm E}^2\,,
\end{equation}
so that the threshold $\delta_{\rm T}$ is reached for 
\begin{eqnarray}
u_{0,{\rm T}}^2 & = & 
\frac{T_{\rm obs}\,v}
{\delta_{\rm T}\,D_{\rm L}}\,
\left(1 - \frac{T_{\rm obs}\,v\,\delta_{\rm T}}{4\,\theta_{\rm E}^2\,D_{\rm L}}\right) 
\nonumber \\
& = & 
\frac{T_{\rm obs}\,v}{\delta_{\rm T}\,D_{\rm L}}\,
\left(1 - F\right)\,,
\end{eqnarray} 
which reveals Eq.~(\ref{varthresh}) for $F \ll 1$. This shows that $F$ measures the asymmetry for $F \ll 1$.

Having found that a significant variation occurs for projected source positions within a circle of
radius $u_{\rm T,var}\,r_{\rm E}$, if $F \ll 1$,
the probability for having an event with a variation of the centroid shift of more than $\delta_{\rm T}$ 
within a given time $T_{\rm obs}$ follows with $\sigma = \pi\,u_{\rm T,var}^2\,r_{\rm E}^2$ from
Eq.~(\ref{probgen}) as
\begin{equation}
\gamma_{\rm var} = \frac{4 \pi G}{c^2}\,D_{\rm S}\, T_{\rm obs}\,
\frac{v}{\delta_{\rm T}}\,
\int\limits_0^1 \rho(x)\,(1-x)\,{\rm d}x\,.
\label{varprob}
\end{equation}
Like the photometric optical depth $\tau_\mu$, 
$\gamma_{\rm var}$ does not depend on the lens masses. 

For a constant mass density $\rho(x) = \rho_0$, one obtains
\begin{eqnarray}
 & & \!\!\!\!\!\!\!\!
\gamma_{{\rm var},0}\;  =\;  \frac{2\pi G}{c^2}\,D_{\rm S}\,T_{\rm obs}\,\frac{v}{\delta_{\rm T}}\,\rho_0 \nonumber \\
& = & 4.3\cdot 10^{-3}\,
\left(\frac{D_{\rm S}}{8.5~\mbox{kpc}}\right)\,
\left(\frac{T_{\rm obs}}{1~\mbox{yr}}\right)\,
\left(\frac{v}{100~\mbox{km s}^{-1}}\right)\,
\times \nonumber \\
 & & \times\,
\left(\frac{\rho_0}{0.08~M_{\sun}\,\mbox{pc}^{-3}}\right)\,
\left(\frac{\delta_{\rm T}}{1~\mbox{$\mu$as}}\right)^{-1}\,.
\label{ratevar}
\end{eqnarray} 
An exponential fall-off of the mass density is discussed in Sect.~\ref{diskscales}.

Values of $\gamma_{\rm var}$ as a function
of the detection threshold $\delta_{\rm T}$  
for bulge sources towards Baade's window and perpendicular
to the Galactic plane are given in Table~\ref{table:gamvar}.

\subsection{Number of events}

\subsubsection{Significant centroid shift}

\begin{table*}
\begin{center}
\caption[]{Rate of events with $\delta > \delta_{\rm T}$} 
\label{table:gampeak}
\begin{tabular}{ccc}
\tableline
\tableline
 & \multicolumn{2}{c}{Rate of events where the centroid shift exceeds the threshold $\delta_{\rm T}$}
 \\
Detection & \multicolumn{2}{c}{per observed star}
 \\
threshold &\multicolumn{2}{c}{-----------------------------------------------------------------------------------------} \\
&  
Bulge stars towards Baade's window & Perpendicular to Galactic plane \\
& $\rho(x) = \rho_0$ & $\rho(x) = \rho_0\,\exp\{-x\,D_{\rm S}/H\}$  \\
& $D_{\rm S} = 8.5~\mbox{kpc}$ & $D_{\rm S} \gg H = 300~\mbox{pc}$ \\
& & \\
$\delta_{\rm T}$ ($\mu$as) & 
$\Gamma_0$ ($\mbox{yr}^{-1}$)& $\Gamma_\infty$ ($\mbox{yr}^{-1}$)\\
\tableline
$1$ & $2.7\cdot 10^{-3}$&$1.9\cdot 10^{-4}$  \\
$5$ & $5.4\cdot 10^{-4}$&$3.8\cdot 10^{-5}$  \\
$10$ &$2.7\cdot 10^{-4}$&$1.9\cdot 10^{-5}$  \\
$100$ & $2.7\cdot 10^{-5}$&$1.9\cdot 10^{-6}$ \\ 
\tableline
\end{tabular}
\tablecomments{The rate of events 
$\Gamma = \gamma_{\rm peak}/T_{\rm obs}
\propto \rho_0\,v\,\delta_{\rm T}^{-1}$   
for which the centroid shift exceeds the
threshold $\delta_{\rm T}$ is shown
for sources (i) towards the Galactic bulge, Eq.~(\ref{ratepeak}), and (ii) perpendicular
to the Galactic plane, Eq.~(\ref{ratevarexp}), with the reference values $v = 100$ km s$^{-1}$,  
and $\rho_0 = 0.08~M_{\sun}\,\mbox{pc}^{-3}$.
}
\end{center}
\end{table*}

Using the criterion $\delta > \delta_{\rm T}$, one can calculate the event rate in analogy
to the photometric case, and count the configurations where
the a source reaches the closest approach to the lens
within the observation time $T_{\rm obs}$ giving rise to a `peak' signature.
As pointed out before, the corresponding area is $\sigma_{\rm peak} = 2\,u_{\rm T}\,r_{\rm E}
\,T_{\rm obs}\,v$.
If one compares this with the area corresponding to
events that show significant variation $\sigma_{\rm var} = \pi\,u_{\rm T,var}^2\,
r_{\rm E}^2$, one sees that
$\sigma_{\rm peak} = (2/\pi)\,\sigma_{\rm var}$, since
$u_{\rm T,var}^2 = T_{\rm obs}\,v/(\delta_{\rm T}\,D_{\rm L})$ and
$u_{\rm T} = \theta_{\rm E}/\delta_{\rm T} = r_{\rm E}/(\delta_{\rm T}\,D_{\rm L})$.
Using the results of the last sections, Eqs.~(\ref{varprob}) and~(\ref{ratevar}),
one obtains a constant event rate
\begin{equation}
\Gamma = \frac{8 G}{c^2}\,D_{\rm S}\,
\frac{v}{\delta_{\rm T}}\,
\int\limits_0^1 \rho(x)\,(1-x)\,{\rm d}x\,,
\end{equation}
and for $\rho(x) = \rho_0$,
\begin{eqnarray}
\Gamma_0 & = & \frac{4 G}{c^2}\,D_{\rm S}\,\frac{v}{\delta_{\rm T}}\,\rho_0 \nonumber \\
& = & 2.7\cdot 10^{-3}\,
\left(\frac{D_{\rm S}}{8.5~\mbox{kpc}}\right)\,
\left(\frac{v}{100~\mbox{km s}^{-1}}\right)\,
\times \nonumber \\
& & \times\,
\left(\frac{\rho_0}{0.08~M_{\sun}\,\mbox{pc}^{-3}}\right)\,
\left(\frac{\delta_{\rm T}}{1~\mbox{$\mu$as}}\right)^{-1}\,\mbox{yr}^{-1}\,.
\label{ratepeak}
\end{eqnarray} 
An exponential fall-off of the mass density is discussed in Sect.~\ref{diskscales}.

Values of $\Gamma$ as a function
of the detection threshold $\delta_{\rm T}$  
for bulge sources towards Baade's window and perpendicular
to the Galactic plane are given in Table~\ref{table:gampeak}.

\subsubsection{Significant variation of centroid shift}

\begin{table*}
\begin{center}
\caption[]{Probability for significant variation and peak}
\label{table:gamvarpeak}
\begin{tabular}{ccc}
\tableline
\tableline
 & \multicolumn{2}{c}{Probability of observing significant variation and peak}
 \\
Detection & \multicolumn{2}{c}{within $T_{\rm obs} = 1~\mbox{yr}$ for a
given observed star}
 \\
threshold &\multicolumn{2}{c}{-----------------------------------------------------------------------------------------} \\
&  
Bulge stars towards Baade's window & Perpendicular to Galactic plane \\
& $\rho(x) = \rho_0$ & $\rho(x) = \rho_0\,\exp\{-x\,D_{\rm S}/H\}$  \\
& $D_{\rm S} = 8.5~\mbox{kpc}$ & $D_{\rm S} \gg H = 300~\mbox{pc}$ \\
& & \\
$\delta_{\rm T}$ ($\mu$as) & 
$\gamma_{{\rm var,peak},0}$ & $\gamma_{{\rm var,peak},\infty}$ \\
\tableline
$1$ & $2.6\cdot 10^{-4}$&$1.4\cdot 10^{-5}$  \\
$5$ & $1.2\cdot 10^{-4}$&$6.3\cdot 10^{-6}$  \\
$10$ & $8.3\cdot 10^{-5}$&$4.4\cdot 10^{-6}$  \\
$100$ & $2.6\cdot 10^{-5}$&$1.4\cdot 10^{-6}$  \\
\tableline
\end{tabular}
\tablecomments{The probability of observing a significant variation and a peak signature in an
event $\gamma_{\rm var,peak} \propto \rho_0\,T_{\rm obs}^{3/2}\,v^{3/2}\,\delta_{\rm T}^{-1/2}$
during
$T_{\rm obs} = 1~\mbox{yr}$ is shown 
for sources (i) towards the Galactic bulge, Eq.~(\ref{probpeakvar}), and (ii) perpendicular
to the Galactic plane, Eq.~(\ref{probpeakvarexp}), with the reference values $v = 100$ km s$^{-1}$,  
$\overline{M^{-1/2}} = (0.5~M_{\sun})^{-1/2}$, and $\rho_0 = 0.08~M_{\sun}\,\mbox{pc}^{-3}$.
}
\end{center}
\end{table*}

As pointed out before, the actual value of the centroid shift is not
measurable, only its temporal variation can be observed. Since it may take
much longer than the observation time to reach a centroid shift smaller than
the detection threshold, there is a difference between whether one considers
$\delta > \delta_{\rm T}$ or the
{\em variation} of $\delta$ larger than $\delta_{\rm T}$. 
Let us consider the probability for a significant variation larger than $\delta_{\rm T}$ {\em and} the
closest approach between lens and source to happen within $T_{\rm obs}$.
Rather than $2\,u_{\rm T}\,r_{\rm E}$, the characteristic width now becomes 
$2\,u_{\rm T,var}\,r_{\rm E}$, and
the area of source positions giving rise to a variation and a peak within $T_{\rm obs}$ is
$\sigma_{\rm var,peak} = 2\,u_{\rm T,var}\,r_{\rm E}\,T_{\rm obs}\,v = 
2\,T_{\rm obs}^{3/2}\,v^{3/2}\,\delta_{\rm T}^{-1/2}\,D_{\rm L}^{-1/2}\,r_{\rm E}$, so that
with Eq.~(\ref{probgen})
\begin{eqnarray}
\gamma_{{\rm var,peak}} & = & 4\,\sqrt{\frac{G}{c^2}}\,D_{\rm S}\,\overline{M^{-1/2}}\,T_{\rm obs}^{3/2}\,
v^{3/2}\,\delta_{\rm T}^{-1/2}\,\times\nonumber \\
&&\times\,\int\limits_0^1 \rho(x)\,\sqrt{1-x}\,{\rm d}x\,,
\label{eq:gammavarpeak}
\end{eqnarray}
where
\begin{equation}
\overline{M^{-1/2}} = \int\limits_0^{\infty} M^{-1/2}\,f_M(M)\,{\rm d}M\,, 
\end{equation}
and for $\rho(x) = \rho_0$ one obtains
\begin{eqnarray}
\gamma_{\rm var,peak,0} & = &\frac{8}{3}\,\sqrt{\frac{G}{c^2}}\,D_{\rm S}\,\overline{M^{-1/2}}\,T_{\rm obs}^{3/2}\,
v^{3/2}\,\delta_{\rm T}^{-1/2}\,\rho_0 \nonumber \\
= & & \!\!\!\!\!\!\!\!\!\!\! 2.6\cdot 10^{-4} \,
\left(\frac{\overline{M^{-1/2}}}{(0.5~M_{\sun})^{-1/2}}\right)\,
\left(\frac{D_{\rm S}}{8.5~\mbox{kpc}}\right)\,
\times \nonumber \\
& & \times\,
\left(\frac{T_{\rm obs}}{1~\mbox{yr}}\right)^{3/2}\,
\left(\frac{v}{100~\mbox{km s}^{-1}}\right)^{3/2}\,
\times \nonumber \\
& & \times\,
\left(\frac{\rho_0}{0.08~M_{\sun}\,\mbox{pc}^{-3}}\right)\,
\left(\frac{\delta_{\rm T}}{1~\mbox{$\mu$as}}\right)^{-1/2}\,.
\label{probpeakvar}
\end{eqnarray}
Note that no constant event rate $\Gamma_{\rm var} = \gamma_{\rm var,peak}/T_{\rm obs}$ is yielded, instead
$\Gamma_{\rm var} \propto T_{\rm obs}^{1/2}$. However, for $T_{\rm obs} \to \infty$, $u_{\rm T,var} \to u_{\rm T}$,
and $\Gamma_{\rm var}$ loses the $T_{\rm obs}$-dependence.

The result for an exponential fall-off of the mass density is discussed in Sect.~\ref{diskscales}.

Values of $\gamma_{\rm var,peak}$ as a function
of the detection threshold $\delta_{\rm T}$  
for sources towards the Galactic bulge and  perpendicular
to the Galactic plane are given in Table~\ref{table:gamvarpeak}.

\section{Measuring the scale parameters of the Galactic disk}
\label{diskscales}
Let us now leave the direction where the mass density is (approximately) constant and
assume a general mass density profile of the form
\begin{equation}
\rho(R,z) = \rho_0\,\exp\left\{-\frac{R-R_0}{d}-\frac{|z|}{h}\right\}\,,
\end{equation}
where $R$ measures the radial position outwards from the Galactic center, $z$ gives the coordinate
perpendicular to the Galactic plane, $R_0$ is the radial position of the sun, $\rho_0$ is the local density
of disk stars, $d$ and $h$ are scale lengths in the Galactic plane and perpendicular to it, where
\begin{equation}
\rho_0 \sim 0.08~M_{\sun}\,\mbox{pc}^{-3}\,, \quad d \sim 3.5~\mbox{kpc}\,, \quad h \sim 0.3~\mbox{kpc}\,.
\end{equation}

For a general direction characterized by the Galactic longitude and latitude $(l,b)$, one has
\begin{equation}
z = x\,D_{\rm S}\,\sin b
\end{equation}
and
\begin{equation}
R = R_0\,\sqrt{1+x^2 y^2 \cos^2 b - 2xy \cos b \cos l}\,,
\end{equation}
where $y = D_{\rm S}/R_0$.
For $b = \pm \pi/2$ (towards the Galactic poles), one obtains $R = R_0$, so that
\begin{equation}
\rho(R,z) = \rho_0\,\exp\left\{-\frac{D_{\rm L}}{h}\right\}\,,
\end{equation}
while for $l = 0$ (towards any
latitude towards the Galactic
center), one obtains $R = R_0 |1-xy \cos b|$, and especially for
for $b=l=0$ (towards the Galactic center), 
$R = |R_0 - x D_{\rm S}|$. For $l=0$ and sources on the same side of the Galactic center as the sun, i.e.
$D_{\rm L}\,\cos b < R_0$, the mass density reads
\begin{eqnarray}
\rho(R,z) & = &	\rho_0\,\exp\left\{-D_{\rm L}\,\left(\frac{|\sin b|}{h}-\frac{\cos b}{d}\right)\right\} \nonumber \\
 & = &	\rho_0\,\exp\left\{-\frac{D_{\rm L}}{H}\right\}\,,
\label{fall2exp}
\end{eqnarray}
where
\begin{equation}
H = \left(\frac{|\sin b|}{h}-\frac{\cos b}{d}\right)^{-1}\,.
\end{equation}
For $b_0^{\pm} = \pm \arctan(h/d) \sim \pm 4.9^{\circ}$, the mass density remains constant as $H \to \infty$, otherwise
the mass density decreases exponentially for $|b| > |b_0|$ or increases 
exponentially for $|b| < |b_0|$
with $D_{\rm L}$ on the length scale $H$, which is equal to $h$ for $b = \pm \pi/2$ and equal to $d$ for
$b = 0$ (increase) or $b = \pi$ (decrease), and a mixture of both scales in general. 

With $s = D_{\rm S}/H$,  
the exponential behavior given by Eq.~(\ref{fall2exp}) can be written in the form
$\rho(x) = \rho_0\,\exp\{-xs\}$, where
$s > 0$ ($H > 0$) means an exponential decrease, $s < 0$ ($H < 0$) 
means an exponential increase,
and $s = 0$ ($|H| \to \infty$) means a constant mass density. 

The expectation value of the lens distance is
yielded with Eqs.~(\ref{taudeltaeq}) and~(\ref{lensdisteq}) as 
\begin{equation} 
<\!\!x\!\!> = \frac{s^2+2s(e^{-s}-2)+6(e^{-s}-1)}{s^3-2s^2+2s(1-e^{-s})}\,.
\end{equation}
For sources at distances $D_{\rm S} \gg H$, one has $s \gg 1$, so that
\begin{equation}
<\!\!x\!\!> = \frac{1}{s}\,,
\end{equation}
which means that 
\begin{equation}
<\!\!D_{\rm L}\!\!> = H\,,
\end{equation}
i.e.\ the expectation value of the lens distance is equal to the scale parameter $H$ of the 
exponential mass distribution.

For a constant mass density along the line-of-sight, 
the optical depth $\tau_{\delta,0}$ is proportional to the source distance $D_{\rm S}$, so that
the optical depth can be written as $\tau_{\delta,0} = \lambda_0\,D_{\rm S}$, where $\lambda_0$ does
not depend on $D_{\rm S}$.
With Eq.~(\ref{taudeltaeq}), the optical depth for an exponential mass density reads
\begin{equation}
\tau_{\delta,s} = 3 \tau_{\delta,0}\,\int\limits_0^1 {\rm e}^{-sx}\,(1-x)^2\,.
\end{equation}
The evaluation of the integral yields
\begin{eqnarray} 
\tau_{\delta,s} & = & 
3 \tau_{\delta,0}\,\left[\frac{1}{s}-\frac{2}{s^2}+\frac{2}{s^3}\left(1-{\rm e}^{-s}
\right)\right] \nonumber \\
& = & 3 \lambda_0\,H\,\left[1-\frac{2}{s}+\frac{2}{s^2}\left(1-{\rm e}^{-s}
\right)\right] \nonumber \\
& =& 3 \lambda_0\,H\,F(s)\,.
\end{eqnarray}
For $s \gg 1$, i.e. $D_{\rm S} \gg H$, and exponential decrease, one obtains
\begin{equation}
F(s) \simeq 1 - \frac{2}{s}\,,
\end{equation}
so that 
\begin{equation}
\tau_{\delta,s} \simeq  
\tau_{\delta,\infty} =  
3 \lambda_0\,H\,,
\label{taudeltavalsexp}
\end{equation}
so that the optical depth measures the scale length $H$. This implies that for different directions,
different combinations of the two disk scale parameters $d$ and $h$ are measured, which means that
with the information from several directions, $d$ and $h$ can be determined.
The case of constant mass density is revealed in the limit $s \to 0$, i.e. $H \to \infty$, where
\begin{equation}
\lim_{s\to 0} \frac{F(s)}{s} = \frac{1}{3}\,,
\end{equation}
so that $\tau_{\delta,s=0} = \tau_{\delta,0}$.
For $s < 0$, the optical depth exceeds $\tau_{\delta,0}$. 
Using $D_{\rm S} = R_0 = 8.5~\mbox{kpc}$ and $d = 3.5~\mbox{kpc}$, one obtains
for the optical depth towards the center of the Galaxy\footnote{Unfortunately, this view is obscured in the
optical}
\begin{equation}
\tau_{\delta,-2.4}(H=d=3.5~\mbox{kpc}) = 2.6\,\tau_{\delta,0}\,(D_{\rm S} = 8.5~\mbox{kpc})\,,
\end{equation}
i.e.\ about 2.5 times larger than towards Baade's window. 

For objects in the LMC ($D_{\rm S} = 50$ kpc), one has approximately $(l,b) = (0,-\pi/2)$, so that
one obtains for $h$ = 0.3~kpc
\begin{equation}
\tau_{\delta,167}(H=h=0.3~\mbox{kpc}) = 0.10\,\tau_{\delta,0}(D_{\rm S} = 8.5~\mbox{kpc})\,,
\end{equation}
while for $h$ = 1 kpc, one obtains
\begin{equation}
\tau_{\delta,50}(H=h=1~\mbox{kpc}) = 0.34\,\tau_{\delta,0}(D_{\rm S} = 8.5~\mbox{kpc})\,.
\end{equation}

Not only the optical depth turns out to be proportional to the scale parameter $H$ for an exponential
fall-off of the mass density and $D_{\rm S} \gg H$, the probabilities for variations, peaks, and 
variation and peaks also share this property.
Like the optical depth,
for constant mass densities,
the probabilities for significant variation $\gamma_{\rm var,0}$, for a peak $\gamma_{\rm peak,0}$, and
for a significant variation and a peak $\gamma_{\rm var,peak,0}$ (Eqs.~(\ref{ratevar}), 
(\ref{ratepeak}), and~(\ref{probpeakvar})) are
proportional to $D_{\rm S}$, so that
$\gamma_{\rm var,0} = \lambda_{\rm var,0}\,D_{\rm S}$,
$\gamma_{\rm peak,0} = \lambda_{\rm peak,0}\,D_{\rm S}$,
and $\gamma_{\rm var,peak,0} = \lambda_{\rm var,peak,0}\,D_{\rm S}$,
where $\lambda_{\rm var,0}$, $\lambda_{\rm peak,0}$, and $\lambda_{\rm var,peak,0}$ do not depend on 
$D_{\rm S}$.

With Eq.~(\ref{varprob}), one obtains for the probability of significant variation for an exponential mass density
\begin{eqnarray}
\gamma_{{\rm var},s} & = & 
2\,\gamma_{{\rm var},0}\,\int\limits_{0}^{1} e^{-sx}\,(1-x)\,{\rm d}x \nonumber \\
& = & 2\,\gamma_{{\rm var},0}\,\left[\frac{1}{s}+\frac{1}{s^2}\left(e^{-s}-1\right)\right] 
\nonumber \\
& = & 2\,\lambda_{{\rm var},0}\,H\left[1+\frac{1}{s}\left(e^{-s}-1\right)\right]\,,
\end{eqnarray}
which yields
for $s \gg 1$
\begin{equation}
\gamma_{{\rm var},s} \simeq
\gamma_{{\rm var},\infty} =
2\,\lambda_{{\rm var},0}\,H\,,
\label{ratevarexp}
\end{equation}
A similar relation holds for $\gamma_{\rm peak}$, since $\gamma_{\rm peak} = (2/\pi)\,\gamma_{\rm var}$.

The probability 
for events to show a peak and significant variation reads with Eq.~(\ref{eq:gammavarpeak}) 
\begin{equation}
\gamma_{{\rm var,peak},s}  =  
\frac{3}{2}\,\gamma_{{\rm var,peak},0}\,
\int\limits_{0}^{1} e^{-sx}\,\sqrt{1-x}\,{\rm d}x\,.
\end{equation}
For $s \gg 1$, the leading order of the integral yields $1/s$,\footnote{Consider e.g. the expansion
of $\sqrt{1-x}$} so that
\begin{equation}
\gamma_{{\rm var,peak},s} \simeq
\gamma_{{\rm var,peak},\infty} =
\frac{3}{2}\,\lambda_{{\rm var,peak},0}\,H\,.
\label{probpeakvarexp}
\end{equation}

\section{The effect of a luminous lens}
\label{blendsection}

\begin{table*}
\begin{center}
\caption[]{The effect of unresolved luminous lens stars}
\label{table:blending}
\begin{tabular}{ccc}
\tableline
\tableline
 & \multicolumn{2}{c}{Correction factor for unresolved luminous lenses}
 \\
Source & \multicolumn{2}{c}{as a function of source magnitude}
 \\
magnitude &\multicolumn{2}{c}{-----------------------------------------------------------------------------------------} \\
&  
Bulge stars towards Baade's window & Perpendicular to Galactic plane \\
& $\rho(x) = \rho_0$ & $\rho(x) = \rho_0\,\exp\{-x\,D_{\rm S}/H\}$  \\
& $D_{\rm S} = 8.5~\mbox{kpc}$ & $D_{\rm S} \gg H = 300~\mbox{pc}$ \\
& & \\
$V_{\rm source}$ & 
$<\!\!(1+g)^{-1}\!\!>$ &
$<\!\!(1+g)^{-1}\!\!>$ \\
\tableline
$12$ & $0.99$ & $0.90$  \\
$15$ & $0.95$ & $0.78$  \\
$17$ & $0.91$ & $0.67$  \\
$19$ & $0.84$ & $0.54$  \\
\tableline
\end{tabular}
\tablecomments{This table shows the effect 
of unresolved luminous lens stars on the number of astrometric microlensing events.
The rate of events with centroid shift $\delta > \delta_{\rm T}$ 
and the probability of observing a significant variation larger than $\delta_{\rm T}$ within the observation time
$T_{\rm obs}$ are decreased by a blending factor $1+g$. The table lists the expectation value $<\!\!(1+g)^{-1}\!\!>$
for several source luminosities and a simple luminosity function as given by Bahcall \& Soneira (\cite{BS}).
Note that there is no dependence on the mass function.}
\end{center}
\end{table*}

For a luminous lens that is not resolved from the source where 
\begin{equation}
g = \frac{L_{\rm L}}{L_{\rm S}}
\end{equation}
is the ratio between the lens and the (unlensed) source apparent luminosities,
one obtains for the magnification (c.f.\ Eq.~(\ref{mueq}))
\begin{equation}
\mu(u) = \frac{g}{1+g} + \frac{u^2+2}{(1+g)\,u \sqrt{u^2+4}}\,,
\end{equation}
which gives for $u \gg 1$
\begin{equation}
\mu(u) = 1 + \frac{2}{(1+g) u^4}\,.
\end{equation}
For the centroid shift relative to
a  source at rest (e.g.\ Boden et al.~\cite{BSV}) one obtains (c.f.\ Eq.~(\ref{deltaeq}))
\begin{equation}
\delta_{\rm S}(u) = \frac{u\,\left(1- g u\sqrt{u^2+4}\right)}{u^2+2+g u \sqrt{u^2+4}}\,\theta_{\rm E}\,.
\end{equation}
However, if one subtracts the proper motion of the apparent `source' object, i.e.\ the centroid of light
composed of source and luminous lens, one obtains 
the observed centroid shift due to lensing as
\begin{eqnarray}
\delta(u) & = & \delta_{\rm S}(u) + \frac{g}{1+g}\,u\,\theta_{\rm E} \\
& = & \frac{u}{1+g}\,\frac{1+g\left(u^2-u\sqrt{u^2+4}+3\right)}{u^2+2+g u \sqrt{u^2+4}}\,\theta_{\rm E}\,,
\end{eqnarray}
which gives for $u \gg 1$
\begin{equation}
\delta(u) \simeq \frac{1}{(1+g)\,u}\,\theta_{\rm E}\,,
\end{equation}
i.e.\ the centroid shift is reduced by a factor $1+g$.
Therefore the threshold for a centroid shift larger than $\delta_{\rm T}$ becomes
\begin{equation}
u_{\rm T}^{\rm blended} = \frac{\theta_{\rm E}}{(1+g)\,\delta_{\rm T}}\,,
\end{equation}
and the threshold for a variation larger than $\delta_{\rm T}$ during the observing time
$T_{\rm obs}$ becomes
\begin{equation}
u_{\rm T,var}^{\rm blended} = \sqrt{\frac{T_{\rm obs}\,v}{(1+g)\,\delta_{\rm T}\,D_{\rm L}}}\,,
\end{equation}
so that in the blended case the detection threshold $\delta_{\rm T}$ is effectively increased
by a factor $1+g$. Therefore, the optical depth $\tau_\delta$ decreases by a factor
$(1+g)^2$, the rate of events where the centroid shift exceeds the threshold and
the probability of a significant variation within $T_{\rm obs}$ decrease
by a factor $1+g$, while the probability of a significant variation and a peak signature
within $T_{\rm obs}$
decreases by a factor $\sqrt{1+g}$.

Since the (disk) lens star is much closer than the source star, one might
think that $g$ is expected to be a large number. However, in a microlensing
experiment, one will only pick the bright source stars, while the lens star
is mostly a faint object. To obtain a more quantitative statement, let us
assume a simple luminosity function for the lenses as given by Bahcall \&
Soneira (\cite{BS}), Eq.~(1), and calculate the expectation value
$<\!\!(1+g)^{-1}\!\!>$ that gives the correction factor for the rate of
events where the centroid shift exceeds $\delta_{\rm T}$ and for the
probability that the centroid shift varies by more than $\delta_{\rm T}$
within $T_{\rm obs}$. The results are shown in Table~\ref{table:blending}.
One sees that the effect is rather small for observing bulge stars towards
Baade's window and somewhat larger for observing perpendicular to the
Galactic plane. In the latter case, the values practically do not depend on
$D_{\rm S}$ if $D_{\rm S} \gg H$. For $V = 17$ sources, the suppression due
to blending is $\sim 10~\%$ towards Baade's window and $\sim 30~\%$
perpendicular to the Galactic plane.

The luminosity function of Bahcall \& Soneira (\cite{BS}) does not take into account a dip around $M_V = 7$ and
a peak around $M_V = 12$ (e.g.\ Kroupa, Tout, \& Gilmore~\cite{KTG}), therefore overestimating the number of stars
around $M_V = 7$ and underestimating the number of stars around $M_V = 12$. However, the values given in
Table~5 depend only weakly on the details of the luminosity function. The most important question about the
luminosity function is up to what point at the low end it remains constant: Bahcall \& Soneira (\cite{BS}) took
it to be constant up to $M_V = 19$ and being zero for $M_V < 19$. A luminosity function that is flat down to
$M_V = 25$ would yield $<\!\!(1+g)^{-1}\!\!> = 0.79$ $(0.70)$ for a $V= 17$ ($V=19$) source in a direction
perpendicular to the Galactic plane, instead
of $<\!\!(1+g)^{-1}\!\!> = 0.67$ $(0.54)$; the values for brighter sources
are less strongly affected.

There is another effect: The formulae given above are valid only if the luminous lens is
{\em not} resolved from the source star. 
If the angular resolution is  $\theta_{\rm res}$, which is 
$\sim$ 200 mas for GAIA and $\sim$ 10 mas for SIM (see, e.g., 
Lindegren \& Perryman~\cite{lind}, 
for more details on GAIA, and B\"oker \& Allen~\cite{boeker},
for more details on SIM), then this limit is reached for 
\begin{eqnarray}
u_{\rm res}  & = & \frac{\theta_{\rm res}}{\theta_{\rm E}} \nonumber \\
& = & 100\,
\left(\frac{\theta_{\rm res}}{200~\mbox{mas}}\right)\,\left(\frac{M}{0.5~M_{\sun}}\right)^{-1/2}\,
\times \nonumber \\
 & & \times\, 
\left(\frac{D_{\rm L}}{1~\mbox{kpc}}\right)^{1/2}\,,
\label{uresexp}
\end{eqnarray}
and lens and source are resolved for $u > u_{\rm res}$. 
This means that the centroid-shift curves
are those for a dark lens in the outer region $u > u_{\rm res}$ and only influenced by
a luminous lens in the inner region $u \leq u_{\rm res}$ irrespective of how large the blend
factor $g$ is.
By comparing $u_{\rm res}$ with the expressions for $u_{\rm T}$, Eq.~(\ref{utexp}), and
$u_{\rm T,var}$, Eq.~(\ref{utvarexp}), one sees that $u_{\rm res}$ is typically smaller than
$u_{\rm T}$ but can be larger or of the order of $u_{\rm T,var}$. Therefore the calculated optical depth
is not strongly
affected by blending, despite the $(1+g)^2$-dependence, because for most of the cases,
the luminous lens is resolved from the source. For the other signatures, the effect of
lens resolution plays a less important role, so that the corresponding probabilities 
are somewhat decreased due to the blending by the unresolved luminous lens. 
Should the angular resolution limit be significantly decreased to, say,
 $\sim 10~\mbox{mas}$ for most of all
discussed cases, the lenses would be resolved and therefore the event rates close to the dark
lens case.

\section{Implications for astrometric space missions}
Upcoming space missions such as SIM and GAIA will provide astrometric
measurements with an accuracy of $\sim$ 4--60~$\mu$as, thus enabling us to observe
the centroid shifts caused by microlensing of stars.

SIM will provide measurements with an accuracy of about $4~\mbox{$\mu$as}$ on targets with
$V < 20$ that
it is pointed to. This will provide the possibility for high-accuracy astrometric
follow-up observations of ongoing microlensing events.
While there is a $\sim 2~\%$ probability that disk stars lead to a centroid shift of the same
order, the variation of this centroid shift during the event duration of the photometric microlensing
event is much smaller, so that the astrometric signal due to the lens that has been
responsible for the original microlensing alert is measured. If one continues to measure the
astrometric signal on larger time scales
$\gtrsim 10~\mbox{yr}$, 
one has to take into account
a contamination due to astrometric microlensing by another lens in the galactic disk.

Contrary to SIM, GAIA will perform an 5-year-long all-sky survey primarily
planned to measure parallaxes with high accuracy (Gilmore et al. 1998) but
does not have the ability of pointing the instrument to a selected target. 
To observe the parallax ellipse, GAIA will perform several measurements on
each target per year. For sources with $V<10$, of which there are about
200,000 objects in the sky, the expected accuracy is $\sim
20~\mbox{$\mu$as}$; for sources with $V<15$, of which there are about 25--35
million objects in the sky, the expected accuracy is $\sim
60~\mbox{$\mu$as}$; and for sources with $V<20$, of which there are about 1
billion objects in the sky, the expected accuracy is $\sim 1~\mbox{mas}$.

Let us now use GAIA as an astrometric microlensing survey instrument and
estimate the expected number of events. Concerning the direction of the
observed stars, let us be conservative with regard to the number of
astrometric microlensing events and consider a direction perpendicular to
the Galactic plane, where the event rate is close to minimum. Let us first
consider the bright ($V < 15$) stars. For an accuracy of $\delta_{\rm T} =
60~\mbox{$\mu$as}$, one estimates with Eq.~(\ref{utexp}) $u_{\rm T} \sim
30$. Therefore, one expects an average event duration (with
Eqs.~(\ref{tEphys})) of
$<\!\!t_{\rm e}\!\!> \sim 3$--$4~\mbox{yr}$. This is smaller than the time of the mission $T_{\rm obs}$,
so that events that reach a certain threshold also vary approximately by the same amount
($u_{\rm T,var} \sim u_{\rm T}$). We can therefore estimate the number of events simply from the
event rate per observed star $\Gamma \sim 3\cdot 10^{-6}/\mbox{yr}$
(Eqs.~(\ref{ratepeak}) and~(\ref{ratevarexp}), Table~\ref{table:gampeak}). Multiplying this with
the 25--35 million stars with $V<15$, the 5 years of the mission and the blending factor of 0.78, one obtains 
about 400~events
during the life-time of GAIA.
For the fainter stars ($V<20$), one obtains for $\delta_{\rm T} = 1~\mbox{mas}$ an event rate
of $\sim 2\cdot 10^{-7}/\mbox{yr}$, so that with 5 years time of the mission, 1 billion stars, and a
blending factor of 0.47, one obtains about 500 events. The very bright stars ($V < 10$) are not expected to
contribute significantly due to their small number.
In total, this estimate gives about 1000 events from the GAIA mission.
We have underestimated this number by the assumption that the mass falls off exponentially 
on a scale of $300~\mbox{pc}$ and
by the assumption that all stars with $V < V_0$ are at $V = V_0$. On the other hand, we have overestimated that
number by the assumption that a signal is detectable when it exceeds the noise threshold (i.e.\ signal-to-noise-ratio
of 1). There is also a dependence on the sampling rate. 

However, we expect the underestimations and the overestimations to cancel to
a big part, so that our estimate indicates the right order. GAIA will thus
observe a large sample of astrometric microlensing events which can be used
to determine the mass and velocities of the lenses, and to determine the
scale length and height of the Galactic disk.

\section{Summary and outlook}

Astrometric and photometric microlensing differ in two main points: First,
the observed centroid shift is a function of both the dimensionless impact
parameter $u$ and the angular Einstein ring radius $\theta_{\rm E}$ such
that for a given $u$, the observed centroid shift is directly proportional
to $\theta_{\rm E}$. On the other hand, the observed magnification is a
dimensionless quantity which depends only on $u$ and not on any other scale.
Second, for large angular separations between the lens and the source, the
centroid shift, being proportional to $1/u$, falls off much more slowly than
the photometric magnitude shift which is proportional to $1/u^4$.  Due to
the dependence of the centroid shift on the angular Einstein radius,
astrometric microlensing favors lenses close to the observer, while
photometric microlensing favors lenses around half-way between observer and
source. Therefore, one gets the largest centroid shifts from nearby objects,
which are the Sun and the planets first, whose effect has to be corrected
for, and then the disk stars. Because of the slower fall-off with the
dimensionless separation $u$ in the astrometric case, detectable signatures
occur for much larger angular separations, so that the average duration of
an event $<\!\!t_{\rm e}\!\!>$ can become much larger than the observation
time $T_{\rm obs}$. For the effect of luminous lenses this means that one
can expect the lens to be resolved from the source star in some of the cases
that show observable signatures. We have shown that the probability that a
disk star introduces a centroid shift larger than a given amount
$\delta_{\rm T}$ at a given time reaches unity for $\delta_{\rm T} 
\sim 0.7~\mbox{$\mu$as}$ for sources towards the Galactic bulge at a latitude
where the mass density of the disk stars is constant, which is a good
approximation for Baade's window, while this probability is about $2\%$ for
$\delta_{\rm T} = 5~\mbox{$\mu$as}$ (see Table 1).  Though there is some
chance that the centroid shift of a photometrically observed microlensing
event, as observed e.g.\ by SIM, is disturbed by disk star lensing (a 2nd
lens), this additional centroid shift is not expected to vary much during
the observation time ($\sim$ several months), so that the effect expected is
a slightly shifted position and the variation of the centroid shift is
determined only by the primary lens.  Only if one extends the observations
to $\sim 10~\mbox{yr}$ after the peak, one has to take the contamination by
disk stars into account.

It is also interesting to examine the expected results from a microlensing
survey looking for centroid shifts rather than the magnification of stars.
As stated earlier, the largest centroid shifts come from nearby objects,
which gives an opportunity to infer information about the disk stars. For
$\delta_{\rm T} \lesssim 10~\mbox{$\mu$as}$ and $T_{\rm obs} \lesssim
10~\mbox{yr}$,
$<\!\!t_{\rm e}\!\!> \gg T_{\rm obs}$. Since one can only measure the variation in the
centroid shift, not its actual value, and since the astrometric
signal does not drop to zero within $T_{\rm obs}$, the condition that the centroid shift
exceeds the threshold $\delta_{\rm T}$ cannot be taken as criterion for an event.
Instead, one has to rely strictly on the criterion that the centroid shift varies by more than
the threshold $\delta_{\rm T}$. For $<\!\!t_{\rm e}\!\!> \ll T_{\rm obs}$, as for most
photometric microlensing events, these two criteria become equivalent.
The probability that a source  star in the 
Galactic bulge towards Baade's window shows a
centroid shift variation larger than $5~\mbox{$\mu$as}$ 
within one year is  $\sim 10^{-3}$, which is about 3 orders
of magnitude larger than the probabilities for photometric microlensing
(see Table~\ref{table:gamvar}). 
Among the events that show significant variations, only a fraction 
(10~\% for $\delta_{\rm T} = 5~\mbox{$\mu$as}$) will have the 
closest angular separation between the lens and the source within the 
observing time, which will result in a clear `peak' signature, namely
an observed change of sign of the
component of the centroid shift parallel to the relative proper motion 
between lens and source, and a maximum of the centroid-shift component 
transverse to it. Since every event `peaks' once, the number of events that reach the
peak within $T_{\rm obs}$ is related to the event rate, while events that show significant
variations only can show this variation in subsequent time intervals.

For an exponential decrease of the mass density along the line of sight (as
it would be the case for lines-of-sight at high Galactic latitudes), the
probabilities for events are proportional to the scale parameter in that
direction if the source stars are at a distance of a few times the scale
parameter or more. For sources perpendicular to the Galactic plane, the
probability for a variation by more than $5~\mbox{$\mu$as}$ and a peak
within
$T_{\rm obs} = 1~\mbox{yr}$ 
is $\sim 6\cdot 10^{-6}$ (Table~\ref{table:gamvarpeak}). By observing astrometric microlensing
events in different directions, one can not only infer information about the total
mass and the mass spectrum but also determine the scale length and scale height of the
Galactic disk.

An advantage of astrometric over photometric observations is that the lens
mass, distance, and velocity can be extracted individually from the
observations (H{\o}g et al.~\cite{Hog}; Miyamoto
\& Yoshii~\cite{miyamoto}; Walker~\cite{walker};
Paczy{\'n}ski~\cite{Pac2}; Boden et al.~\cite{BSV}).

We expect $\sim 1000$ astrometric microlensing events to be detected
by the GAIA mission during its lifetime of 5 years.

\acknowledgements{Martin Dominik's work at STScI has been financed by
research grant Do 629/1-1 from Deutsche Forschungsgemeinschaft, while his
work in Groningen is financed by a Marie Curie Fellowship from the European
Commission (ERBFMBICT972457). It is a pleasure to thank Ron Allen, Torsten
B\"oker, Stefano Casertano, Erik H{\o}g, Mario Lattanzi, Michael Perryman,
and Jayadev Rajagopal for helpful discussions regarding the details of the
SIM and the GAIA missions.}

\end{document}